\documentclass[numberedappendix]{emulateapj}
\begin{document}

\slugcomment{Accepted for Publication in the Astronomical Journal: July 18, 2010}


\title{Structure and Composition of Two Transitional Circumstellar Disks in 
Corona Australis}

\shorttitle{CrA Transition Disks}
\shortauthors{Hughes et al.}

\author{
A.~M.~Hughes\altaffilmark{1}, 
S.~M.~Andrews\altaffilmark{1,2}, 
D.~J.~Wilner\altaffilmark{1}, 
M.~R.~Meyer\altaffilmark{3}, 
J.~M.~Carpenter\altaffilmark{4}, 
C.~Qi\altaffilmark{1}, 
A.~S.~Hales\altaffilmark{5}, 
S.~Casassus\altaffilmark{6}, 
M.~R.~Hogerheijde\altaffilmark{7}, 
E.~E.~Mamajek\altaffilmark{8}, 
S.~Wolf\altaffilmark{9}, 
T.~Henning\altaffilmark{10}, 
M.~D.~Silverstone\altaffilmark{11}}

\email{mhughes@cfa.harvard.edu}

\altaffiltext{1}{Harvard-Smithsonian Center for Astrophysics,
60 Garden Street, Cambridge, MA 02138; mhughes, sandrews, dwilner, 
cqi$@$cfa.harvard.edu}
\altaffiltext{2}{Hubble Fellow}
\altaffiltext{3}{Institute for Astronomy ETH, Physics Department, HIT
J 22.4, CH-8093 Zurich, Switzerland; mmeyer$@$phys.ethz.ch}
\altaffiltext{4}{California Institute of Technology, Department of Astronomy, 
MC 105-24, Pasadena, CA 91125, USA; jmc$@$astro.caltech.edu}
\altaffiltext{5}{National Radio Astronomy Observatory, 520 Edgemont Road, 
Charlottesville, Virginia, 22903-2475, USA; ahales$@$alma.cl}
\altaffiltext{6}{Departamento de Astronomía, Universidad de Chile, Casilla 
36-D, Santiago, Chile; simon$@$das.uchile.cl}
\altaffiltext{7}{Leiden Observatory, Leiden University, P.O. Box 9513, 2300 RA,
  Leiden, The Netherlands; michiel$@$strw.leidenuniv.nl}
\altaffiltext{8}{University of Rochester, Department of Physics \& Astronomy, Rochester, NY 14627-0171, USA; emamajek$@$pas.rochester.edu}
\altaffiltext{9}{University of Kiel, Institute of Theoretical Physics and Astrophysics, Leibnizstrasse 15, 24098 Kiel, Germany; wolf$@$astrophysik.uni-kiel.de}
\altaffiltext{10}{Max-Planck Institute for Astronomy, Koenigstuhl 17, D-69117, Heidelberg, Germany; henning$@$mpia-hd.mpg.de}
\altaffiltext{11}{Eureka Scientific, Inc., 113 Castlefern Dr., Cary, NC 25713, USA; mdsilverstone@att.net}

\bibliographystyle{apj}

\begin{abstract}

The late stages of evolution of the primordial circumstellar disks surrounding
young stars are poorly understood, yet vital to constrain theories of planet
formation.  We consider basic structural models for the disks around two 
$\sim$10\,Myr-old members of the nearby RCrA association, RX~J1842.9-3532 
and RX~J1852.3-3700.  We present new arcsecond-resolution maps of their 
230\,GHz continuum emission from the Submillimeter Array and unresolved 
CO(3-2) spectra from the Atacama Submillimeter Telescope Experiment.  By 
combining these data with broadband fluxes from the literature and infrared 
fluxes and spectra from the catalog of the Formation and Evolution of 
Planetary Systems (FEPS) Legacy program on the {\em Spitzer} Space Telescope, 
we assemble a multiwavelength data set probing the gas and dust disks.  Using 
the Monte Carlo radiative transfer code  \texttt{RADMC} to model 
simultaneously the SED and millimeter continuum visibilities, we 
derive basic dust disk properties and identify an inner cavity of radius 
16\,AU in the disk around RX~J1852.3-3700.  We also identify an optically thin 
5\,AU cavity in the disk around RX~J1842.9-3532, with a small amount of 
optically thick material close to the star.  The molecular line observations 
suggest an intermediate disk inclination in RX~J1842.9-3532, consistent with
the continuum emission.  In combination with the dust models, the molecular 
data allow us to derive a lower CO content than expected, suggesting that the 
process of gas clearing is likely underway in both systems, perhaps
simultaneously with planet formation.  

\end{abstract}

\section{Introduction}

One of the key problems in planet formation is understanding how the reservoir 
of planet-forming material -- the disk of gas and dust around a young star -- 
evolves with time.  Perhaps the most compelling stage in the evolution of 
protoplanetary disks is the ``transitional'' stage during which gas and dust 
are cleared from the system \citep[e.g.][]{str89,skr90}.  This stage appears 
to be either rapid or rare, since fewer than 10\% of low- to intermediate-mass 
stars in young star-forming regions are typically observed to be transitional 
systems \citep[e.g.][]{sil06,cie07,uzp08}, perhaps with some dependence on
stellar mass \citep[e.g.][]{cur10}.  Transition disks are traditionally 
identified empirically by a deficit of mid-infrared dust excess over the 
stellar photosphere relative to comparable young systems.  This
deficit is associated with a lack of hot dust close to the star.  The advent 
of the {\it Spitzer} Space Telescope has revolutionized the quality and 
quantity of available data on the dust content of young stellar systems, 
particularly transitional objects \citep[see, e.g.,][]{naj07,cur09,muz10}.  
One of its many contributions has been to reveal a new class of gapped or 
``pre-transitional'' disks, in which an anomalously faint mid-infrared 
spectrum is bracketed by substantial near- and far-infrared excesses 
\citep{esp07,esp08,esp10}.  Follow-up of {\it Spitzer}-identified transitional 
systems with high spatial resolution observations of continuum emission at 
millimeter wavelengths has led to the confirmation that mid-IR spectral 
deficits are associated with a lack of long-wavelength emission from the disk 
center \citep{cal02,cal05,hug07,hug09,bro07,bro08,bro09,pie07,esp08}.  The 
properties of systems with gaps and holes are beginning to provide valuable 
insight into the physical mechanism(s) responsible for the dispersal of the 
circumstellar disk.  Photophoresis, the presence of unknown binary companions, 
grain growth, dynamical interactions with giant planets in formation, and 
photoevaporation have all been suggested as clearing mechanisms; there is 
some indication that different processes may dominate at different ages 
\citep[see, e.g.,][]{sic10}. 

The Formation and Evolution of Planetary Systems (FEPS) Legacy program on the
{\it Spitzer} Space Telescope \citep{mey06,car08} is a spectrophotometric 
survey of nearby sun-like stars, with masses from 0.7 to 1.3 M$_\sun$ 
and ages between 3 Myr and 3 Gyr.  These ages bracket the period of time when 
gas and dust were cleared from the primordial Solar nebula, and the epoch when 
the Solar system achieved its present configuration.  For 309 objects in the 
FEPS sample, the survey includes IRAC 3.6-8.0\,$\mu$m 
photometry to probe for hot, dusty analogues to the asteroid belt in the Solar 
system, MIPS 24 and 70\,$\mu$m photometry to probe dust in the Kuiper Belt 
regions, and IRS 5-40\,$\mu$m spectra to search for mineralogical features.  
The excellent coverage of the infrared spectral energy distribution (SED) 
permits modeling of the temperature, size, composition, and an initial 
estimate of the spatial distribution of dust grains \citep[see, 
e.g.,][]{kim05,bou08, cor09}.  The IRS spectra are particularly useful for 
identifying systems with inner cavities or gaps in their dust distribution.  
However, there are many degeneracies inherent in the derivation of spatial 
information from unresolved spectra, and the SED provides little information 
about the large grains that comprise most of the dust mass in the system.  
It is therefore desirable to combine the information from the SED with 
spatially resolved observations at millimeter wavelengths.  Spectral line 
observations of low-level rotational emission from the CO molecule can provide 
a complementary probe of the molecular gas content, which is the dominant mass 
constituent, and can yield important clues to the evolution of transitional 
objects. 

In this paper, we analyze the SEDs and resolved millimeter continuum emission
of two sources from the FEPS sample, RX~J1842.9-3532 and RX~J1852.3-3700.  
These sources were detected in the {\it ROSAT} All-Sky Survey and identified 
as young stars by \citet{neu00}.  They have spectral type K2 and K3, 
respectively \citep{car08}, and have been classified as classical T Tauri 
stars (cTTSs) based on the presence of strong H$\alpha$ emission lines.  
Neither was identified as a multiple-star system in the speckle-interferometric 
observations of \citet{koh08}.  They are located within a few degrees of the 
CrA molecular cloud \citep[distance 130\,pc;][]{neu00}, and have kinematics and 
secular parallaxes consistent with the RCrA association (E. Mamajek, private 
communication).  The dust grain composition of the inner disks of both systems 
was modeled in detail by \citet{bou08}, and \citet{pas07} report the detection 
of [Ne II] emission likely arising from a hot disk atmosphere and calculate 
accretion rates for both systems of order 10$^{-9}$\,M$_\sun$\,yr$^{-1}$.  
The estimated stellar ages are $\sim$10\,Myr, among the oldest in the 
1-10\,Myr range for T~Tauri stars in CrA measured by \citet{neu00}.  These 
sources were selected for study on the basis of their age, their brightness 
in the 1.2\,mm continuum \citep{car05}, and their accessibility to the 
Submillimeter Array (SMA)\footnote{The Submillimeter Array is a joint project 
between the Smithsonian Astrophysical Observatory and the Academia Sinica 
Institute of Astronomy and Astrophysics and is funded by the Smithsonian 
Institution and the Academia Sinica.}, so that the spectral information from 
the FEPS survey could be combined with resolved submillimeter observations.  
They were also chosen for their proximity on the sky, which increases the 
efficiency of submillimeter observations by allowing them to share 
calibrators.  The high-quality {\it Spitzer} IRS spectra provide constraints 
on the dust composition and temperature structure on the two systems, and 
both exhibit a flux deficit in the mid-IR photospheric excess that points to 
the presence of an inner hole or gap.

Sections~\ref{sec:obs} and \ref{sec:results} describe the collection of new 
data from the SMA and the Atacama Submillimeter Telescope Experiment (ASTE) 
that complement the spectra and broadband fluxes from the FEPS survey and 
the literature.  In Section~\ref{sec:fit} we present the tools and 
techniques that we use to model the SED and resolved millimeter-wavelength 
data, and we present the models of the dust disk structures in 
Section~\ref{sec:dust_model}.  In Section~\ref{sec:CO_model}, we explore the
dust disk model in the context of the constraints on the gas content of the
two systems.  We summarize our results and discuss their implications in 
Section~\ref{sec:discussion}. 

\section{Observations and Data Reduction}
\label{sec:obs}

\subsection{SMA Observations}

The SMA observations of the two sources took place on 2005 May 14 during a
full six-hour track with six of the 6-meter diameter antennas operating in the 
compact-north configuration, yielding baseline lengths between 10 and 180 
meters (8 and 140\,k$\lambda$).  The phase stability was adequate for 
most of the track, with phase differences of 20-30 degrees between calibrator 
scans, but the phases lost coherence during the last hour of the night.  The 
weather was fair, with the 225\,GHz atmospheric opacity increasing from 0.10 
to 0.14 throughout the night.  Observations of the two FEPS sources were 
alternated with observations of the quasar J1924-292 at 15-minute intervals 
to calibrate the atmospheric and instrumental variations of phase and 
amplitude gain.  Callisto was used as the flux calibrator, with a calculated 
brightness of 6.59 Jy; the derived flux of J1924-292 was 5.4\,Jy, with an 
estimated uncertainty of $\lesssim 20$\%.  The correlator was configured to 
provide a spectral resolution of 512 channels over the 104 MHz\,bandwidth in
the chunk containing the CO(2-1) line, corresponding to a velocity resolution 
of 0.26\,km\,s$^{-1}$.  Two other 104\,MHz chunks were observed at high 
resolution at frequencies corresponding to the $^{13}$CO(2-1) and C$^{18}$O(2-1)
lines.  The remainder of bandwidth in each 2\,GHz-wide sideband was devoted to 
measuring the 230\,GHz continuum, observed at a spectral resolution of 
4.2\,km\,s$^{-1}$.  The data were calibrated using the MIR software 
package\footnote{http://cfa-www.harvard.edu/$\sim$cqi/mircook.html}, 
and imaging was carried out with the MIRIAD software package.

\subsection{ASTE Observations}

Observations of both FEPS sources took place on 2008 June 28 and 29 
using the CATS345 receiver on the 10.4-meter ASTE dish.  RX~J1842.9-3532
was observed on both nights, while RX~J1852.3-3700 was observed only on the
second night.  The receiver was tuned to place the CO(3-2) rotational 
transition in the lower sideband with the HCO$^+$(4-3) transition in the 
upper sideband.  The high resolution spectrometer mode was used to partition 
the 128\,MHz bandwidth into 1024 channels, yielding a spectral resolution of 
0.11\,km/s.  Position switching was used to subtract the instrumental and sky 
background.  In order to ensure that the detected CO emission originated at 
the position of the star, we used an offset position 1.5\,arcmin to the 
east on the night of June 28 and 1.5\,arcmin to the west on June 29 
and averaged the baseline-subtracted spectra to create the final spectrum.  
The double-peaked CO(3-2) line from RX~J1842.9-3532 is detected independently 
on both nights using the different offset positions, which allows us to 
localize the emission to within 1.5 arcmin of the star ($\sim$4 beam widths).  

The CO(3-2) and HCO$^+$(4-3) tuning of the receiver resulted
in a spurious 1.11\,MHz sinusoidal ripple of variable amplitude and phase 
across the bandpass, which was subtracted individually from each 10-second 
integration in the following manner.  The amplitude and frequency of the ripple 
were estimated by finding the peak in a fast fourier transform of the spectrum,
and then a least squares fit was performed to determine the precise amplitude,
phase, and frequency of the sinusoidal ripple, plus the slope and intercept of
a linear component to remove the worst of the baseline features.  During this 
process, the region of the spectrum containing the line was not 
included in the fit so as to avoid inadvertently subtracting it.  
Integrations with an abnormally large ripple amplitude or highly irregular 
baseline shape across the bandpass were discarded (roughly 10\% of each data
set).  After subtracting the sinusoid and linear fit, a third-order polynomial 
was subtracted from each integration using the CLASS software package\footnote{
http://www.iram.fr/IRAMFR/GILDAS}, fitting the 20\,km\,s$^{-1}$ to each side 
of the region that appeared to contain the line.  The 10-second integrations 
were then averaged together to produce a spectrum for each night; the 
spectra for each night were averaged and weighted according to their total 
integration time ($\sqrt{1/t_\mathrm{int}}$) to produce the final spectrum.   
Due to differing exposure times and poorer weather on the night of June 29, 
the rms noise in the spectrum of RX~J1842.9-3532 is 24\,mK, while the rms 
noise in the spectrum of RX~J1852.3-3700 is 39\,mK.  To derive the absolute 
flux scale, we calculated the main beam efficiency using observations of the 
calibrator M17SW taken throughout the night.  Assuming a peak main beam 
temperature in the CO(3-2) line of 85.3K, derived on the CSO 10.4\,m telescope 
by \citet{wan94}, we derive main beam efficiencies that varied between 0.55 
and 0.63 over the course of the two nights.

\section{Results}
\label{sec:results}

\subsection{Millimeter Continuum}
\label{sec:continuum}

\begin{figure*}
\epsscale{0.9}
\plotone{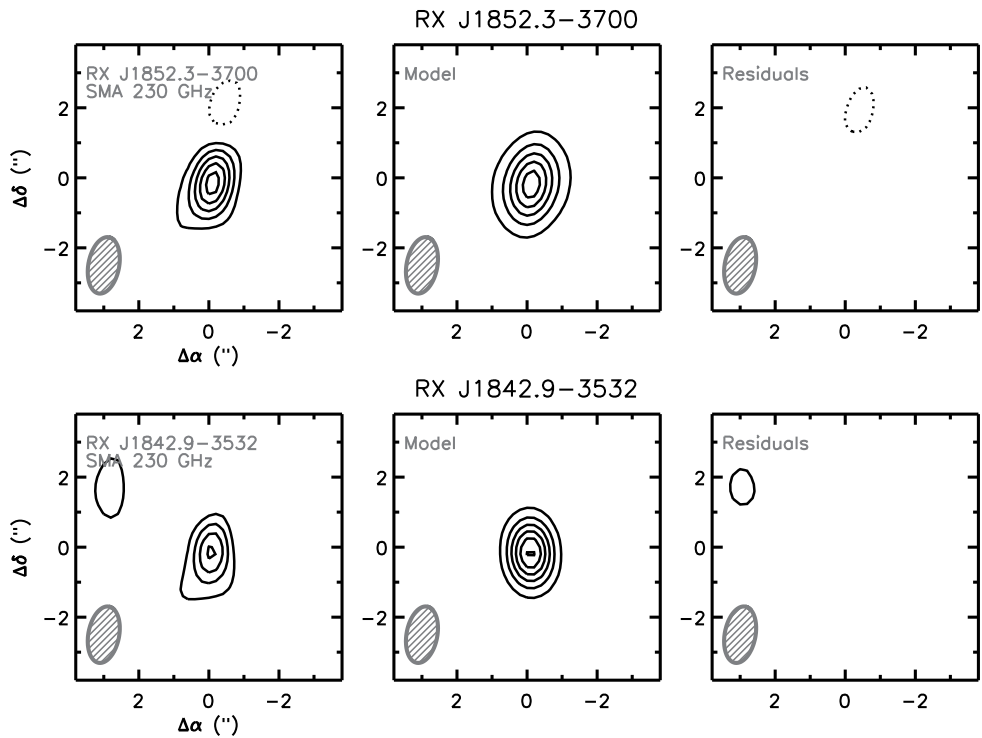}
\figcaption{SMA 230 GHz maps of the continuum emission from RX~J1852.3-3700
({\it top}) and RX~J1842.9-3532 ({\it bottom}).  Each row shows three panels:
the SMA data ({\it left}), the representative model ({\it center}), and the
residuals ({\it right}).  The contour levels are 
[2,4,6,...]$\times$3.5\,mJy\,beam$^{-1}$ (the rms noise), with solid lines 
indicating positive contours and dotted line indicating negative contours.  
The 1\farcs0$\times$1\farcs7 synthesized beam is indicated by the ellipse in 
the lower left corner.  
\label{fig:cont_map}
}
\end{figure*}

With the SMA observations, we detect 230\,GHz continuum emission from the 
outer disks around both target stars.  The contour maps in the left panels
of Figure \ref{fig:cont_map} show the strong detection of emission centered 
on the star position at the middle of the field.  The spatial resolution of 
the data is 1\farcs0$\times$1\farcs7, or $\sim$150\,AU at a distance of 
130\,pc, well matched to the typical size of a circumstellar disk but 
insufficient to provide evidence for the presence or absence of an inner 
cavity.  In order to estimate the integrated flux and approximate outer disk 
geometry, a Gaussian fit to the visibilities was performed using the MIRIAD 
task \texttt{uvfit}.  For RX~J1842.9-3532, the fit yields an integrated flux 
of 49$\pm$8\,mJy and indicates that the disk is only marginally resolved: 
the major and minor axes lengths of 0\farcs74$\pm$0\farcs32 and 
0\farcs44$\pm$0\farcs32 suggest that the disk is probably not viewed face-on,
but do not place strong constraints on the inclination.  We estimate an 
inclination angle of 54$^\circ$ based on these measurements, which is 
consistent with the morphology of the ASTE CO(3-2) spectrum described in 
Section \ref{sec:CO_obs} below.  The fit to the RX~J1852.3-3700 visibilities 
yields a flux of 60$\pm$8\,mJy and major and minor axes of length 
0\farcs76$\pm$0\farcs21 and 0\farcs73$\pm$0\farcs21, consistent with a nearly 
face-on geometry; we use these measurements to estimate an inclination of 
16$^\circ$.  The inclination estimates are highly uncertain, but the 
intermediate geometry of RX~J1842.9-3532 is supported by the line profile in 
Section \ref{sec:CO_obs} and the more face-on geometry of RX~J1852.3-3700 is 
supported by the H$\alpha$ line profile modeling of \citet{pas07}.  While 
observations at higher resolution would be advantageous for constraining the 
detailed mass distribution, even the rudimentary estimates of disk geometry 
provided by these observations are useful for constraining the disk properties 
when combined with constraints from the broadband SED.  Simultaneous modeling 
of the SED and millimeter-wavelength visibilities is described in Section 
\ref{sec:fit} below.

\subsection{CO(2-1) and CO(3-2) Line Observations}
\label{sec:CO_obs}

We do not detect molecular gas emission from either system in the 
interferometric SMA observations of the CO(2-1), $^{13}$CO(2-1), or 
C$^{18}$O(2-1) lines.  The data provide a 3$\sigma$ upper limit of 
0.4\,Jy\,beam$^{-1}$ in each 0.26\,km\,s$^{-1}$ channel, with a synthesized 
beam size of 1\farcs5$\times$0\farcs8.  Although the disks are only marginally 
resolved in the continuum emission, there is reason to expect that the extent 
of CO(2-1) emission may be several times larger than that of the continuum 
\citep[see, e.g.,][]{hug08}.  As a result, spatial filtering may be a factor 
in the non-detection (see further discussion in Section~\ref{sec:CO_model}).  

\begin{figure}[t]
\epsscale{1.0}
\plotone{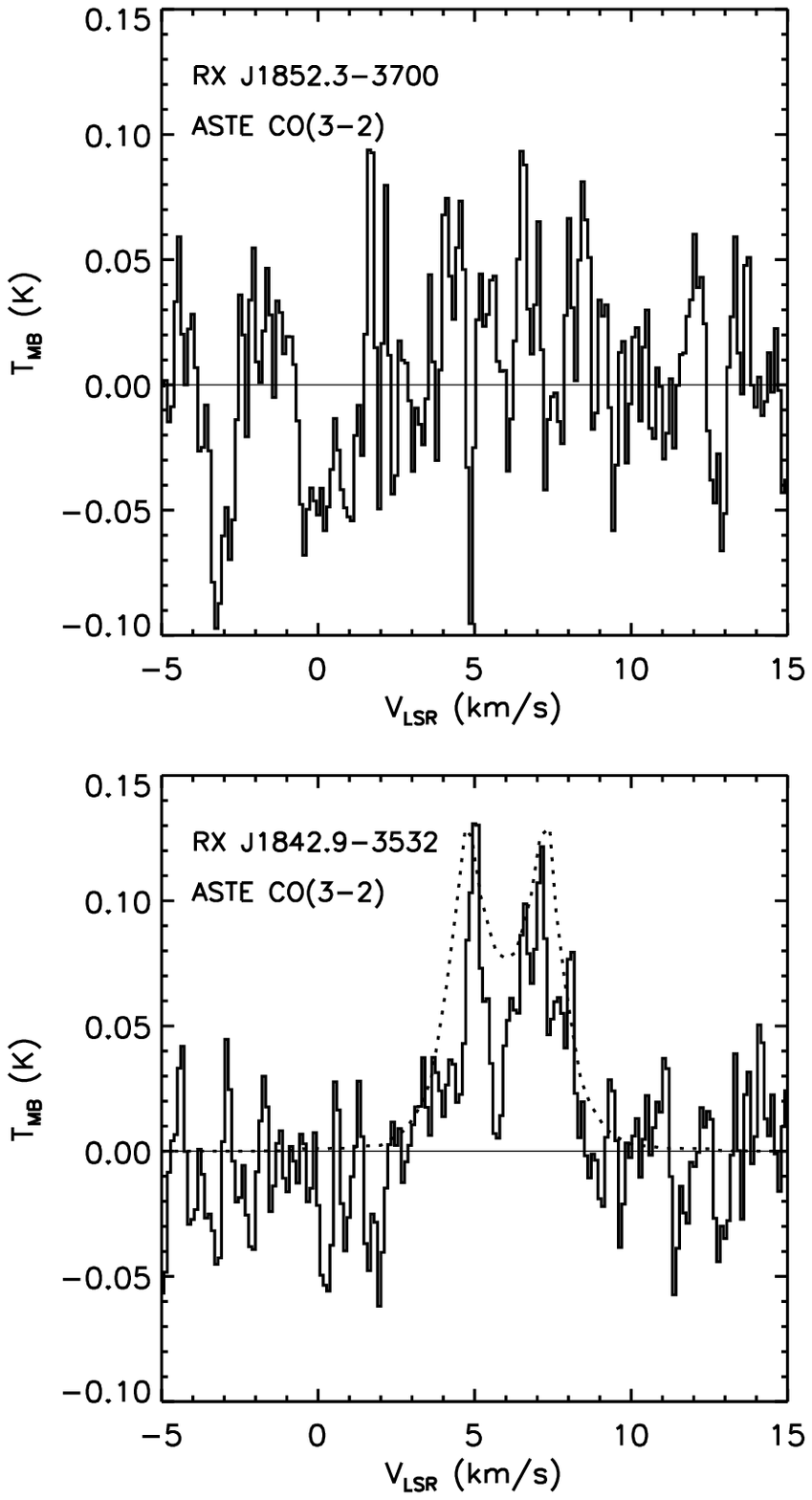}
\figcaption{ASTE CO(3-2) spectra of the disks around RX~J1852.3-3700 
({\it upper}) and RX~J1842.9-3532 ({\it lower}).  No emission is detected
from the RX~J1852.3-3700 system.  The RX~J1842.9-3532 emission (solid line) 
displays the characteristic double-peaked profile characteristic of an 
inclined structure in Keplerian rotation about the central star.  The 
line profile predicted by the SED- and visibility-based model of the dust 
disk structure (dotted line) compares favorably with the observations.
\label{fig:spec}
}
\end{figure}

We do not detect any CO(3-2) emission in the ASTE observations of the disk 
around RX~J1852.3-3700, with an rms of 39\,mK in each 0.1\,km\,s$^{-1}$ 
channel.  Observations of the disk around  RX~J1842.9-3532 reveal a 
double-peaked line profile, shown as a solid black line in 
Figure \ref{fig:spec}.  The integrated strength of the CO(3-2) line is 
0.24\,K\,km\,s$^{-1}$ with a peak main-beam brightness temperature of 130\,mK 
and FWHM of 2.6\,km\,s$^{-1}$.  The double-peaked profile is consistent with 
material in Keplerian rotation about the star, viewed at an intermediate 
inclination angle of $\sim$54$^\circ$.  We detect no emission from the CrA
molecular cloud near the line in velocity space, although it is possible that 
absorption from the cloud in the vicinity of the disk might influence the line 
shape despite the sources' large distance from CrA cloud center.  In Section 
\ref{sec:CO_model} below, we investigate the relationship of the CO(3-2) 
emission to the dust properties, including implications for the disk geometry 
and gas-to-dust mass ratio.

\section{Analysis}

\begin{figure*}[t]
\epsscale{1.0}
\plotone{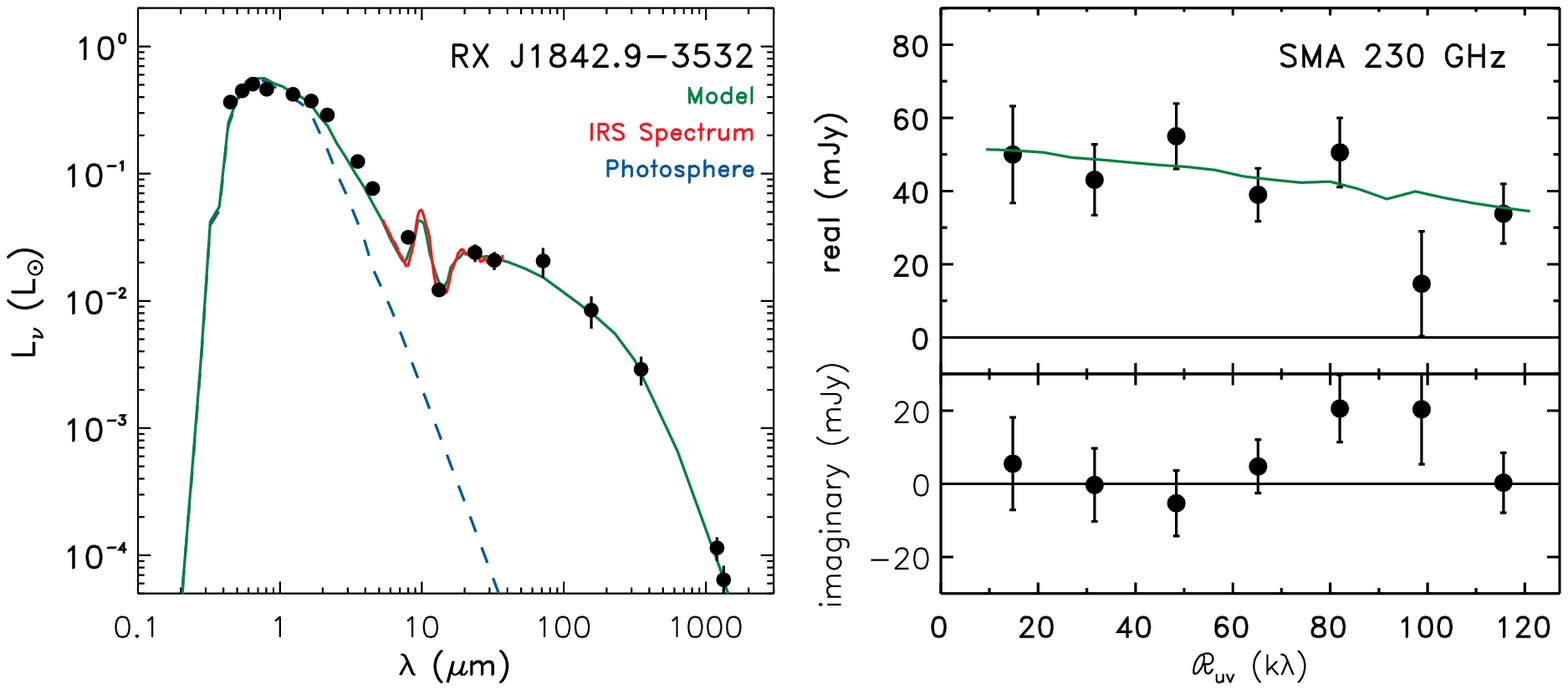}
\figcaption{Spectral energy distribution ({\it left}) and the real and 
imaginary components of the deprojected SMA 230\,GHz visibilities ({\it 
right}) for RX~J1842.9-3532.  The broad-band SED (black points) and IRS
spectrum (red line) are well reproduced by the best-fit \texttt{RADMC} disk 
structure model (green line).  The model stellar photosphere (dashed blue 
line) is plotted for comparison.  The units of the ordinate are defined so 
that $\mathrm{L}_\nu = 4 \pi d^2 \nu \mathrm{F}_\nu$ in units of L$_\sun$.  
For a mathematical definition of the abscissa, refer to \citet{hug08}; the
deprojection is carried out as in \citet{lay97}.
\label{fig:rxj1842}
}
\end{figure*}

\begin{figure*}[t]
\epsscale{1.0}
\plotone{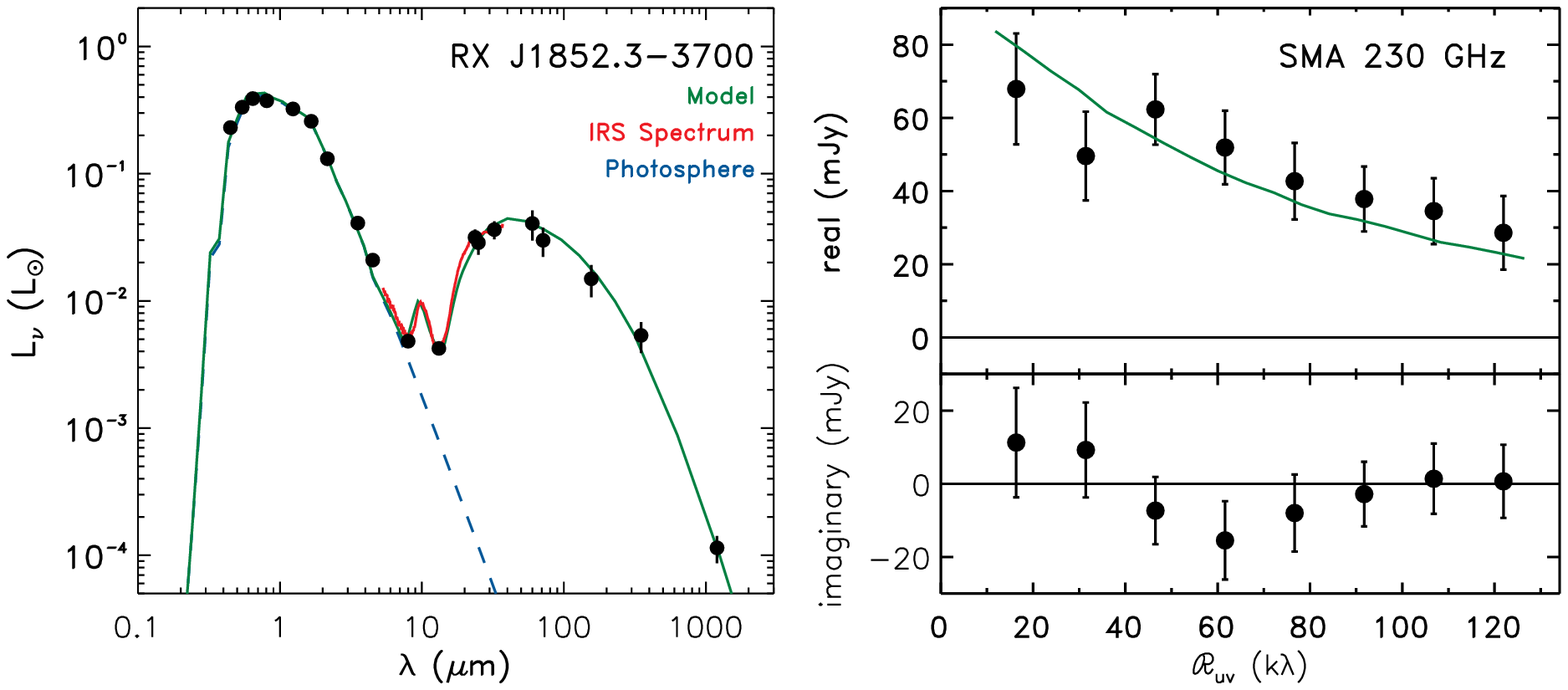}
\figcaption{Same as Fig.~\ref{fig:rxj1842} above, but for RX~J1852.3-3700.  
\label{fig:rxj1852}
}
\end{figure*}

In order to characterize the basic properties of the disks, we seek 
a model that can reproduce the observational features of each
system.  We assemble a data set that combines the millimeter-wavelength 
properties of the gas and dust described above with constraints from the 
broadband SED and IRS spectrum.  We use the IRS spectrum and 
SED from the FEPS database \citep[described in][]{car08}, including 
optical, near-IR, and millimeter fluxes collected from the literature 
\citep{neu00,skr06,car05}.  Figures~\ref{fig:rxj1842} and \ref{fig:rxj1852} 
show the SED (black points) and the IRS spectrum (red line) in the 
left panel for each disk, alongside the SMA 230\,GHz visibilities (black 
points) in the right panel.  In order to improve the signal-to-noise ratio 
of the plotted data, the visibilities have been deprojected \citep[see, 
e.g.,][]{lay97} according to the disk geometry inferred in 
Section~\ref{sec:continuum} and averaged in bins of 15\,k$\lambda$.  For a 
mathematical description of the abscissae of the visibility plots, refer to 
\citet{hug08}.  

\subsection{Modeling the SED and Millimeter Visibilities}
\label{sec:fit}

In an effort to reproduce these observations, we generated synthetic
broadband SEDs, {\it Spitzer} IRS spectra, and millimeter continuum
visibilities using the radiative transfer method and disk structure models
described by \citet{and09a}.  In these flared, axisymmetric disk structure
models, the radial surface density profile is characterized by a
similarity solution for viscous accretion disks, $\Sigma \propto 
(R_c/R)^{\gamma} \exp{[-(R/R_c)^{2-\gamma}]}$, where $R_c$ is a characteristic 
radius and the normalization is proportional to the disk mass \citep[for 
simplicity, the radial index has been fixed to $\gamma = 1$;][]{lyn74,har98}.  
Vertically, the densities are distributed as a Gaussian with a scale height 
that varies as a power-law with radius, $H \propto R^{1+\psi}$, with scale
height $H_{100}$ at a radius of 100\,AU.  This parametric definition of the 
vertical dust distribution is maintained to mimic the sedimentation of dust 
grains below the disk atmosphere \citep[e.g.,][]{dul04b}; no attempt is made 
to iterate on the density structure to force the dust into vertical 
hydrostatic equilibrium.  To model the cleared inner disks for these 
transitional sources, we scale down the surface densities by a factor 
$\delta_{\Sigma}$ inside a radius $R_{\rm cav}$ \citep[$\Sigma_{\rm cav} = 
\delta_{\Sigma} \Sigma$;][]{and09a}.  Moreover, in an effort to better 
reproduce the detailed shape and solid state features in the IRS spectra, we 
permit a small (multiplicative) increase in the scale-height at the cavity edge
($\delta_H$) and adjust the dust grain properties in the inner disk
\citep[for details, see][]{and09b}.  Inner disk silicate opacities are taken
from the Jena database \citep[see, e.g.,][]{oss94}. 

For a given parametric disk structure, fixed input stellar information
\citep{car08}, and opacities \citep[see][]{and09a}, we use the
two-dimensional Monte Carlo radiative transfer code \texttt{RADMC}
\citep{dul04a} to calculate an internally-consistent temperature structure
and generate synthetic data products that can be compared to the
observations.  However, the parameter degeneracies introduced by the
additional inner disk parameters and the high quality of the IRS spectra
(requiring attention to mineralogy) make the minimization method described 
by \citet{and09a} prohibitive.  Instead, we aimed to find a representative 
model that can reproduce the basic features of the data by focusing 
on varying parameters like the cavity size ($R_{\rm cav}$) and surface density 
reduction ($\delta_{\Sigma}$).  These models serve as initial estimates of 
the disk structures that can be refined when future observations are available 
(e.g., high angular resolution millimeter data).

\subsection{Representative Models}
\label{sec:dust_model}

Table~\ref{tab:model} presents the parameters of representative disk structure 
models capable of reproducing the observational data for both systems, and 
indicates those parameters that were fixed by particular observational 
constraints.  The 130\,pc distance to the RCrA association is from 
\citet{neu00} and the visual extinction and stellar properties are drawn from 
the FEPS database \citep{car08}, while the inclination and position angle 
are estimated from the data as described in Sections~\ref{sec:continuum} and 
\ref{sec:CO_obs} above.  The other parameters are defined in 
Section~\ref{sec:fit}; \citet{and09a,and09b} include extensive discussion of 
the degeneracies between parameters and the ways in which the observational 
features are linked to the components of the disk structure model.  Here we 
include comments on several parameters that are particularly relevant for 
reproducing the data described in this paper. 

{\it Inner disk structure ($R_{cav}$ and $\delta_{cav}$)} --- 
These parameters are tied primarily to the wavelength and magnitude of the 
rise in the infrared flux longward of the 10\,$\mu$m silicate feature.  It 
should be noted that while the density reduction $\delta_\mathrm{cav}$ is 
greater for RX~J1842.9-3532 than for RX~J1852.3-3700, the initial difference 
in surface density must be taken into account: because RX~J1842.9-3532 is 
almost a factor of four more compact than RX~J1852.3-3700, the surface density 
throughout most of the disk, including within the cavity, is larger.  As a 
result, the inner disk of RX~J1852.3-3700 is entirely optically thin, while 
that of RX~J1842.9-3532 includes both optically thick and optically thin 
regimes.  The surface density profile of the two models is plotted in 
Figure~\ref{fig:sigma}.  As indicated in Section~\ref{sec:discussion}
below, the details of the inner disk structure are not well constrained by
these models, although the presence of an inner cavity of greatly reduced
surface density is firmly indicated.

{\it Puffing at inner edge of outer disk ($\delta H$)} ---
The parameter $\delta \mathrm{H}$, introduced in \citet{and09b}, describes the 
extent to which the scale height at the edge of the cavity is puffed up, which 
occurs because of the frontal illumination of the inner edge of the outer 
disk and is primarily manifested observationally in the shape of the far-IR 
SED.  While a small $\delta \mathrm{H}$ can help to account for the very 
steep mid- to far-IR jump in flux observed in the RX~J1852.3-3700 IRS 
spectrum, no shadowing is required to reproduce the spectrum of 
RX~J1842.9-3532.  

{\it Inner disk dust properties} ---
We have very little information about the composition of dust grains in
the outer disk, other than that millimeter-size grains must be present to 
account for the emission at long wavelengths.  However, reproducing the
features in the IRS spectrum requires us to vary the composition of the hot
grains in the inner disk and wall that give rise to these features, as in 
\citet{and09b}.  The 10\,$\mu$m silicate feature and steep rise in flux near 
20\,$\mu$m from the disk around RX~J1852.3-3700 are reproduced well by an 
inner disk and cavity edge populated entirely by small ($\sim$0.1\,$\mu$m) 
amorphous silicate grains.  For the disk around RX~J1842.9-3532, the strength 
and position in wavelength of the silicate feature are well reproduced by a 
mixture of small ($a_\mathrm{max}$=1\,$\mu$m) and large ($a_\mathrm{max}$=1\,mm)
ISM-composition grains ($\sim$80\%) and crystalline and amorphous silicates 
($\sim$20\%) in the inner disk.  This combination of grain compositions is by 
no means a unique solution to the problem of fitting the mid-IR spectrum, but 
merely demonstrates that a mixture of different grain properties is helpful in 
accounting for the observed spectral features.  A more detailed mineralogical 
analysis of these systems can be found in \citet{bou08}.  The estimated mass
of dust in the inner disk is $10^{-10}$\,M$_\sun$ for RX~J1842.9-3532 and
$10^{-6}$\,M$_\sun$ for RX~J1852.3-3700.  

The model SED and millimeter visibilities for the structural parameters in
Table~\ref{tab:model} are shown by the green lines in Figures~\ref{fig:rxj1842} 
and \ref{fig:rxj1852}.  They reproduce the basic features of all of the 
available dust disk diagnostics, including the broadband SED, the IRS 
spectrum, and the millimeter-wavelength visibilities.  In the discussion 
below, we focus on the most robustly-constrained model parameters, including
the extent and surface density reduction of the inner cavity and the size 
and dust mass of the outer disk.

\begin{table}[ht]
\caption{Star and Estimated Disk Parameters}
\begin{tabular}{lcc}
\hline
Parameter & RX~J1842.9-3532 & RX~J1852.3-3700 \\
\hline
\multicolumn{3}{c}{Fixed} \\
\hline
Distance (pc) & 130 & 130 \\
A$_{\mathrm V}$ (magnitudes) & 1.06 & 0.97 \\
Spectral Type & K2 & K3 \\
Log $g$ (log\,cm\,s$^{-2}$) & 4.27 & 4.31 \\
T$_\mathrm{eff}$ (K) & 4645 & 4854 \\
$i$ ($^\circ$) & 54 & 16 \\
P.A. ($^\circ$) & 32 & -56 \\
$\gamma$ & 1.0 & 1.0 \\
\hline
\multicolumn{3}{c}{Varied} \\
\hline
R$_{\mathrm C}$ (AU) & 50 & 180 \\
M$_{\mathrm D}$ (M$_\sun$)$^a$ & 0.010 & 0.016 \\
$\psi$ & 0.2 & 0.2 \\
H$_{100}$ (AU) & 4.8 & 6.3 \\
R$_\mathrm{cav}$ (AU) & 5 & 16 \\
$\delta_\mathrm{cav}$ & $9\times10^{-6}$ & $3\times10^{-6}$ \\
$\delta \mathrm{H}$ & 1 & 1.4 \\
\hline
\label{tab:model}
\end{tabular}
\tablenotetext{a}{Total mass in gas and dust, assuming a gas-to-dust mass
ratio of 100}
\end{table}

\begin{figure*}[t]
\plotone{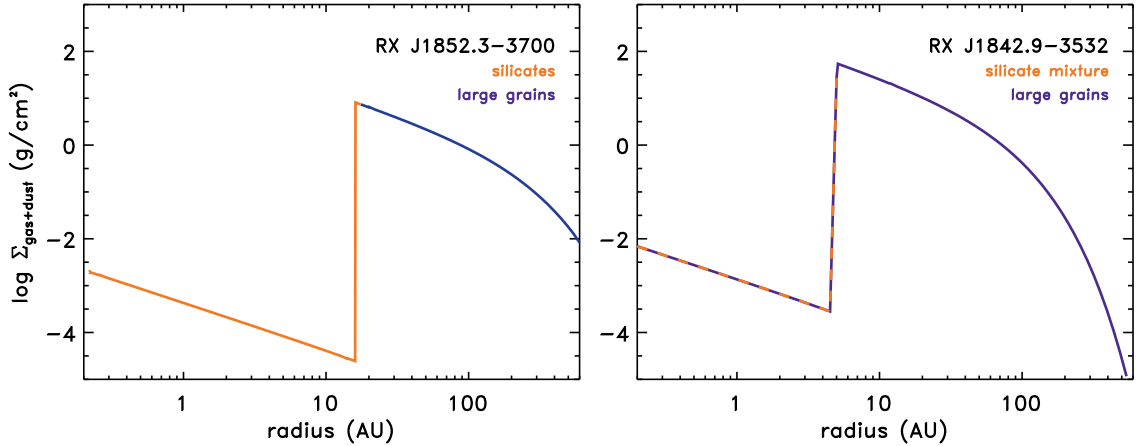}
\figcaption{
Surface density profiles for the representative model parameters in 
Table~\ref{tab:model}.  The line colors indicate the dust grain composition
at each position within the disk; the dust grain composition is described in
Section~\ref{sec:dust_model}.  The surface density incorporates the total 
mass in gas and dust, assuming a gas-to-dust mass ratio of 100. 
\label{fig:sigma}
}
\end{figure*}

\subsection{Constraints on Molecular Gas Content}
\label{sec:CO_model}

Here we compare the predictions of the dust disk model with the constraints 
on the CO emission described in Section~\ref{sec:CO_obs}.  For simplicity, 
we assume that gas and dust are well-mixed, with identical spatial and 
temperature distributions.  We use the standard assumptions of a gas-to-dust   
mass ratio of 100:1 and a CO abundance of 10$^{-4}$ relative to H$_2$.  To
account for the depletion of gas-phase CO onto dust grains at low temperatures,
we decrease the CO abundance by an additional factor of 10$^{-4}$ in regions 
with temperatures $\le$20\,K.  As in \citet{and09a}, we then use the Monte 
Carlo molecular line radiative transfer code \texttt{RATRAN} \citep{hog00} 
to calculate the level populations and predict the sky-projected intensity 
of CO arising from each system, given the underlying structure of the 
representative models derived in Section~\ref{sec:dust_model}.  We use the 
\texttt{MIRIAD} task \texttt{convol} to convolve the resulting intensity 
distributions with the 21\farcs1 beam of the 10.4\,m ASTE telescope, since the 
ASTE spectra provide the most stringent limits on the CO emission from the 
systems.  

With the standard assumptions of gas-to-dust and CO-to-H$_2$ ratios, 
the model projection of CO(3-2) line flux strongly overpredicts both 
the upper limit for the disk around RX~J1852.3-3700 and the detection of 
CO(3-2) emission from the disk around RX~J1842.9-3532 by a factor of 4.8.  
In order for the model to successfully reproduce the weak emission from 
RX~J1842.9-3532, the volume density of CO must be decreased to 8$\pm$3\% of 
its initial value.  A model spectrum for this case is given by the dotted 
line in Figure~\ref{fig:spec}, and compares well with the strength and width 
of the observed CO spectrum.  While the line peaks appear narrower than the 
model, the noise in the line is too large to merit modeling the profile in 
detail; it is also possible that contamination from remnant molecular cloud 
material could contribute to the narrowing of the peaks.

It is worthwhile to consider whether the apparently faint CO emission from
these disks could be produced by overestimated model gas temperatures,
rather than low CO abundances.  In our models, we have assumed that the
gas and dust are well-mixed.  However, the infrared emission from these
sources is too faint to accomodate a dust distribution in vertical
hydrostatic equilibrium: the dust has apparently settled toward the disk
midplane.  The gas, on the other hand, should still feel a vertical pressure 
gradient, and therefore have a larger scale-height than the dust 
\citep[e.g.,][]{dul04b,dal06}.  As a result the gas in our model, which is 
assumed to follow the spatial distribution of the dust, is at somewhat lower 
temperatures than it would be in a thicker disk.  This difference in temperature
due to the physical location of the gas and dust is apparently exacerbated 
by exposure to high-energy (FUV, X-ray) radiation from the central star 
\citep{qi06}.  Simulations show that these heating processes typically
raise the gas temperature over that of the dust in the upper disk layers from
which the CO emission arises \citep[e.g.,][]{kam04,woi09}.  If anything, our 
model assumptions should then bias the results toward {\it underestimating} 
gas temperatures in the disk atmospheres, and therefore under-predict the 
amount of CO emission.

If instead we take the well-mixed models at face value (i.e., assuming 
identical gas and dust temperatures), the model gas temperatures can only be 
modified by adjusting the dust structure parameters.  The model temperatures 
in the outer disk where the CO emission is generated are primarily set by 
the size of the disk (i.e., distance from the star), and the amount of flaring. 
To reduce the gas temperatures by the factor of $\sim$5 required to reproduce 
the observations for standard CO/H$_2$ and gas-to-dust conversions, we would 
need to increase the emitting area ($R_c$) by roughly an order of magnitude 
or decrease the scale-height ($H$) to an extreme, flat-disk level: these 
possibilities are inconsistent with the millimeter visibilities and infrared 
SED, respectively.  Given these difficulties, a more straightforward 
interpretation of the faint line emission in these disks is that the CO 
abundances are substantially lower than would be expected for the standard 
gas-to-dust mass ratio of 100 and/or CO abundance relative to H$_2$ of 
10$^{-4}$ \citep[see also][]{qi04,hug09}.

In the absence of measurements of the H$_2$ content of the outer disk, it is 
not possible to determine whether the lower CO content results from a 
reduction in gas-to-dust mass ratio or abundance of CO relative to H$_2$, but 
in either case it marks a significant departure from standard assumptions.  
Because the rms in the ASTE spectrum of RX~J1852.3-3700 is larger than that 
of RX~J1842.9-3532, the upper limit on the CO content of the disk is similarly 
$\sim$8\% of the initial value, assuming a standard gas-to-dust ratio and CO 
abundance.  The model with reduced CO content relative to standard assumptions 
is also consistent with the limits on CO(2-1) emission from the SMA.  We use
RATRAN to generate a sky-projected CO(2-1) emission map, which is then sampled
with the fourier components of the SMA data using the MIRIAD task 
\texttt{uvmodel} to account for spatial filtering effects.  The model with
standard CO abundance should be detected by the SMA observations, whereas the
model with 8\% CO content is consistent with the upper limits on the CO(2-1)
emission from both systems.

If the gas disk were truncated relative to the dust disk, this could contribute
to the low CO content.  It is unlikely, however, that both systems would undergo
truncation -- especially given the dearth of companions within 6'' 
\citep{koh08} -- and the truncation would have to be severe in order to 
account for an order of magnitude reduction in CO content.  It should also be 
noted that the conclusion of reduced CO content is largely independent of the 
model parameters describing the inner disk and the transitional region between 
inner and outer disk.  The CO(3-2) emission arises only from the cold outer 
disk, and the gas-to-dust ratio is derived only for this region, since we 
have no information on the gas content of the inner disks of these systems.  
The extent of the outer disk and its total dust mass are derived from two 
observational parameters: the millimeter flux and size scale indicated by 
the resolved visibilities.  While the vertical structure and inner disk 
properties can affect the temperature of the outer disk and therefore the 
magnitude of the CO(3-2) emission, these effects are secondary to the basic 
midplane temperature structure determined by the radial scale of the dust 
disk.  The reduced CO content is therefore robust to variations in the details 
of the inner disk structure, since variations in inner disk properties will 
have only second-order effects on the gross outer disk properties from which 
this conclusion is derived. 

Both systems therefore appear to have undergone a reduction in molecular gas 
content relative to the standard assumptions for primordial disks.  Given 
their age and transitional SEDs, this may indicate that gas dispersal is 
underway simultaneously with dust clearing from the inner disk.

\section{Discussion}
\label{sec:discussion}

We have generated models that can reproduce simultaneously the basic 
observational diagnostics of the gas and dust disks around RX~J1842.9-3532 
and RX~J1852.3-3700, including their broadband SEDs, IRS spectra, resolved 
millimeter-wavelength visibilities, and CO(3-2) spectra.  As indicated by the 
mid-IR flux deficit, both systems are transitional, with an inner cavity of 
significantly decreased dust optical depth.  

The disk around RX~J1842.9-3532 also exhibits a substantial near-IR excess 
over the stellar photosphere.  It shares this feature with the sample of
objects labeled gapped, or ``pre-transitional'' by \citet{esp07}.  They model
such systems with an optically thin inner disk bracketed by an optically thick 
ring close to the star and the optically thick outer disk at large radii.  
Similar models for the LkCa 15 system, refined with the addition of radiative 
transfer through the inner disk, are described in \citet{mul10}.
\citet{ise09} modeled the near-IR excess and mid-IR deficit in the LkCa 15 
system using a density distribution that increases with distance from the 
star, but includes a puffed-up inner rim at the dust disk edge.  In our study, 
the inner disk model retains the continuous surface density profile of the 
outer disk (decreasing with distance from the star), suppressed by the factor 
$\delta_\mathrm{cav}$ (see Figure~\ref{fig:sigma}), with no change in scale 
height at the inner edge of the inner disk.  Due to the relatively small scale 
heights in the inner disk, we can approximate the optical depth to starlight 
as the product of surface density and 1\,$\mu$m opacity, $\Sigma_\mathrm{R} 
\kappa_{1 \mu \mathrm{m}}$.  In this approximation, the cavity is optically 
thick between 0.01 and $\sim$0.2\,AU but optically thin between $\sim$0.2 and 
5\,AU, comparable to the models described in \citet{esp07}.  These results 
suggest that transition disks with near-IR excess are not necessarily 
``gapped'' in terms of their surface density or discontinuous in terms of their
scale height, since we demonstrate that the inner disk can be modeled using 
a single, continuous surface density function for the disk cavity that 
contains just enough mass to have both optically thick and optically thin 
regimes.  Effectively, this indicates that we can place no constraint on the 
contrast in surface density or scale height between the ``gap'' and the 
optically thick ring near the star based on the morphology of the IRS 
spectrum.  This is reflected by the success of several very different models 
of inner disk structure \citep[][this work]{esp07,ise09,mul10} in reproducing 
the characteristic mid-IR deficit surrounded by near- and far-IR excesses.

The model parameters for the two disks discussed in this paper are consistent 
with those of the nine disks in Ophiuchus that were studied using this 
method by \citet{and09a}.  This is perhaps unsurprising, since these targets 
were similarly selected on the basis of their large submillimeter fluxes.  
RX~J1842.9-3532 and RX~J1852.3-3700 have slightly lower masses due to 
missing material in the otherwise dense disk center, as for the transitional 
systems in the high-resolution Ophiuchus sample.  Yet as with the other 
transition disks in \citet{and09a} they are still on the high end of 
the distribution of masses of Taurus and Ophiuchus disks in the sample of 
\citet{and05,and07}.  The transition disk frequency for very low-mass stars in 
the CrA region has been examined by \citet{sic08,erc09}, with estimates ranging
from 15-50\%.  On the whole, the disks around RX~J1842.9-3532 and 
RX~J1852.3-3700 appear comparable to transition disks in other low-mass star
forming regions.

While relatively little is known about the gas evolution of circumstellar
disks, it is somewhat surprising that such massive dusty disks should have such
low CO content.  It is interesting to note that other transitional systems 
around K stars have also been found to have low CO content through similar 
modeling approaches to that presented here: TW Hya \citep{qi04} and GM Aur 
\citep{hug09} have derived CO abundances of $10^{-5}$ and $10^{-6}$, 
respectively \citep[see also][]{kas97}.  The low gas content of TW Hya has 
recently been corroborated by Herschel observations that appear to indicate 
a gas-to-dust mass ratio of $\sim$2.6-26, substantially lower than the 
standard value of 100 derived for the ISM \citep{thi10}.  In the absence of 
knowledge about the H$_2$ content of the system, this is consistent with the 
\citet{qi04} measurement of low CO abundance, which assumed a gas-to-dust mass 
ratio of 100.  To be clear, the inferred CO content of these systems is 
judged to be ``low'' in comparison with the standard assumptions of a 
gas-to-dust mass ratio of 100 and a CO abundance relative to H$_2$ of 
10$^{-4}$.  A reduction in either total molecular gas content (i.e., gas 
dispersal) or CO relative to H$_2$ (e.g., molecular depletion) could be 
responsible, or the standard assumptions could be incorrect.  While apparently 
consistent with ISM conditions, there are almost no observational constraints 
on either gas-to-dust or CO-to-H$_2$ ratios in the outer regions of primordial 
(non-transitional) protoplanetary disks.  It is particularly important to 
understand depletion, since molecular depletion onto dust grains has been 
suggested as a possible cause of low molecular gas content in circumstellar 
disks \citep[see, e.g.,][]{dut97}.  Depletion is sensitive to the details of 
the temperature and density structure of the disk, which may be difficult to 
determine at low spatial resolution.  The high resolution survey of 
\citet{and09a} finds CO abundances consistent with standard assumptions to 
within a factor of a few for the three Ophiuchus disks exhibiting strong CO 
emission, although they note that most of the sources in the sample are 
substantially contaminated by the local molecular cloud within which the 
stars are embedded.  Comparable modeling efforts for solar-type systems less 
embedded in their natal clouds are therefore highly desirable for determining 
the implications of the low CO content of these transitional systems.  

Assuming that the standard assumptions are correct for protoplanetary disks, 
it is interesting to consider the implications of low CO content for the 
physical processes clearing the cavities in transitional systems.  Among 
the several mechanisms proposed to clear the central regions of disks, the 
most popularly invoked are gravitational interaction with a giant planet in 
formation \citep[e.g.,][]{lin86,bry99} and photoevaporation by energetic 
radiation from the star \citep[e.g.,][]{cla01,ale06}.  Many of the dust
properties of transitional systems, including the disk masses inferred from 
dust flux alone and the sharp transition between the inner and outer disk from
SED modeling, point towards planet formation as the most likely process.
The mass accretion rates of the observed systems, particularly in combination
with the disk masses from dust, are also generally more consistent with 
planet formation as the dominant physical mechanism \citep[see, 
e.g., discussions in][]{naj07,ale07}.  RX~J1852.3-3700 in particular occupies
an interesting and underpopulated region of the $\dot{M}$-$M_*$ plane 
investigated by \citet{naj07}, due to its extremely low accretion rate measurement.  The identification of gapped or 
``pre-transitional'' systems like RX~J1842.9-3532 has also been interpreted
as being more consistent with planet formation than photoevaporative clearing 
\citep[e.g.,][]{esp07}.  It is worth noting, however, that a reduced gas mass 
for transitional systems relative to protoplanetary disks would present some
difficulty for the planet formation scenario, since there is no reason {\it 
a priori} to expect the formation of giant planets to alter the molecular gas 
content of the outer disk.  The masses in Table \ref{tab:model} may be 
misleading, since they represent the total mass in gas and dust of the 
\texttt{RADMC} model assuming a standard gas-to-dust mass ratio of 100, 
without taking into account the evidence for low CO content described in 
Section~\ref{sec:CO_model}; if the low CO content is a result of reduced 
gas-to-dust mass ratio, the disk masses may be reduced by an order of 
magnitude or more, placing them squarely within the region of parameter 
space preferred by photoevaporative clearing models.  If this were generally
true of transition disks (including TW Hya and GM Aur), it could also provide
an alternative explanation for the trend in mass accretion rate as a function
of disk mass noted by \citet{naj07}, since an order of magnitude difference
in total mass between the protoplanetary and transition disks would bring the
two populations into agreement.  

Photoevaporation of disk material by energetic stellar radiation can similarly
account for many, but not all, features of these systems.  Its most obvious
success is in explaining the low CO content of the outer disk, which cannot
easily be accomplished within the framework of planet formation.  
One complicating factor, however, is the measured accretion rate of material
onto the stars based on modeling of the H$\alpha$ profiles by \citet{pas07}. 
They calculate an accretion rate of $1\times10^{-9}$\,M$_\sun$\,yr$^{-1}$ for 
RX~J1842.9-3532 and $5\times10^{-10}$\,M$_\sun$\,yr$^{-1}$ for RX~J1852.3-3700. 
These are roughly an order of magnitude below the average for 1\,Myr-old stars 
in Taurus \citep{gul98,cal04}, which is consistent with the trend for 
transitional systems in Taurus noted by \citet{naj07}.  RX~J1852.3-3700 in 
particular occupies an underpopulated region of the $\dot{M}$-$M_*$ plane 
investigated by \citet{naj07}, due to its extremely low accretion rate 
measurement, which could indicate the presence of a very massive planet in
the disk.  This relatively low but measurable accretion rate is inconsistent 
with the original predictions of most photoevaporation models \citep[see, 
e.g.,][]{ale07}.  Recent work by \citet{owe10} suggests that accretion rates 
of this magnitude may in fact be consistent with the predictions of 
radiation-hydrodynamic models that incorporate both x-ray and EUV 
photoevaporation; however, they assume a stellar x-ray luminosity far greater 
than is typically observed, and EUV luminosities of young stars are not yet 
well determined.  Another potential problem is that the 10$^5$\,yr timescales 
predicted for photoevaporative clearing is only marginally consistent with 
a $\sim$10\% transition disk fraction at ages of 10$^{6}$-10$^7$\,yr.  
Certainly more observations are needed to understand the physical mechanisms 
responsible for clearing the inner cavities of transition disks, including 
observations across a larger sample and a range of ages, to determine the 
molecular gas content of transitional relative to primordial disks and 
constrain the processes responsible for their dissipation. 

At least two well-studied systems at ages of $\sim$10\,Myr have been 
identified as candidate transition disks undergoing photoevaporation: the
M star HD~100453 \citep{col09} and the A star 49 Ceti \citep{hug08b}.  The 
former exhibits a strong IR excess indicative of an optically thick outer 
disk, but with no evidence of accretion and a maximum gas-to-dust ratio of 
4:1 in the outer disk.  The latter exhibits dust properties similar to a 
debris disk, yet retains an extended optically thin molecular gas disk with 
an inner hole.  The differences between these systems and the CrA transition 
disks in this paper are striking, and are perhaps indicative of the range of 
evolutionary paths over which the transition from protoplanetary to debris 
disk may occur. 

\section{Summary and Conclusions}

The 10\,Myr old systems RX~J1842.9-3532 and RX~J1852.3-3700 present a rare 
opportunity to study the late-stage evolution of circumstellar disks.  We
assemble a multiwavelength data set including the broadband SED, {\it Spitzer} 
IRS spectrum, and generate models of the disk structure for each system that 
can reproduce the features of the dust continuum emission.  We infer the
presence of an optically thin 16\,AU cavity in the disk around RX~J1852.3-3700.
The near-IR photospheric excess and mid-IR deficit in the SED of 
RX~J1842.9-3532 allow us to infer the presence of an optically thin 4\,AU 
gap in the disk around RX~J1842.9-3532, bracketed by an optically thick ring 
near the central star ($\sim$0.01-0.2\,AU) and an optically thick outer disk.  
This configuration marks it as a member of the recently identified class of
``pre-transitional'' objects \citep{esp07}.  Comparison of the dust disk 
models with spatially unresolved CO(3-2) spectra from ASTE allow us to 
infer a molecular gas content roughly an order of magnitude less than that
expected for standard assumptions about the gas-to-dust mass ratio and 
CO abundance relative to H$_2$.  If the standard assumptions are correct
for primordial disks, then this low CO content is suggestive that evolution
of the gas disk may be underway simultaneously with the dust disk dispersal.

The transitional systems in CrA observed in this paper therefore mark 
interesting test cases for distinguishing between proposed mechanisms for
gas and dust clearing at late ages.  Follow-up of these objects with 
instruments at various wavelengths can help to fill in our picture of the 
properties of the inner and outer disks.  For example, the far superior 
spatial resolution of the Atacama Large Millimeter Array (ALMA) will permit 
vastly improved modeling of the structure of the extended gas and dust disk, 
as well as providing direct access to the conditions within the cavity, 
removing ambiguity about surface densities and scale heights in the inner 
disk.  The sensitivity to spectral line emission provided by ALMA and {\it 
Herschel} will yield insight into the gas mass and chemistry and therefore 
the origin of the reduced CO content of the outer disk.  In the meantime, 
observations of rovibrational lines can aid in determining the gas content 
of the warm inner disk, which will aid in distinguishing between proposed 
clearing mechanisms.  Scattered light images would also be useful for 
constraining the vertical structure of the disks and reducing degeneracies
in these initial models.  The suite of instruments currently coming online 
is poised to revolutionize our ability to characterize the physics of 
individual disks in the compelling transitional stage of evolution.

\acknowledgements
We thank C.~P. Dullemond for access to the 2-dimensional version of RADMC.  
Partial support for this work was provided by NASA 
Origins of Solar Systems Program Grant NAG5-11777.  A.~M.~H. acknowledges 
support from a National Science Foundation Graduate Research Fellowship.  
Support for S.~M.~A. was provided by NASA through Hubble Fellowship grant 
\#HF-01203-A awarded by the Space Telescope Science Institute, which is 
operated by the Association of Universities for Research in Astronomy, Inc., 
for NASA, under contract NAS 5-26555.  M.~R.~M. acknowledges the Harvard 
Origins of Life Initiative, the Smithsonian Astrophysical Observatory, and a 
NASA TPF Foundation Science Program grant NNG06GH25G (PI: S. Kenyon) for 
sabbatical support.

\bibliography{ms}

\end{document}